\documentclass[a4paper,fleqn]{cas-dc}

\usepackage[numbers]{natbib}
\usepackage{lipsum}
\usepackage{balance}
\usepackage{multirow}
\usepackage{soul} 
\usepackage{color}
\usepackage{caption}
\usepackage{subcaption}
\usepackage{graphicx}
\usepackage{listings}
\usepackage{blindtext}
\usepackage[algoruled,linesnumbered,lined]{algorithm2e}
\usepackage{enumitem} 
\usepackage{url}
\usepackage{fancyhdr}
\usepackage{booktabs}
\usepackage{colortbl}
\usepackage{etoolbox}
\usepackage{verbatimbox,xcolor}
\usepackage{algpseudocode}
\usepackage{algorithmicx}
\usepackage{ragged2e}
\usepackage{hyperref}
\usepackage[yyyymmdd]{datetime}
\usepackage{accents}
\usepackage[most]{tcolorbox}
\usepackage{fontawesome5}
\usepackage[normalem]{ulem}
\usepackage{multirow}
\usepackage{soul} 
\usepackage{color}
\usepackage{caption}
\usepackage{graphicx}
\usepackage{listings}
\usepackage[most]{tcolorbox}

\renewcommand{\ttdefault}{txtt}
\definecolor{anti-flashwhite}{rgb}{0.95, 0.95, 0.96}
\definecolor{lightgray}{rgb}{0.83, 0.83, 0.83}
\definecolor{dkgreen}{rgb}{0,0.6,0}
\definecolor{gray}{rgb}{0.5,0.5,0.5}
\definecolor{mauve}{rgb}{0.58,0,0.82}
\definecolor{light-gray}{gray}{0.80}
\definecolor{gray}{rgb}{0.4,0.4,0.4}
\definecolor{darkblue}{rgb}{0.0,0.0,0.6}
\definecolor{cyan}{rgb}{0.0,0.6,0.6}

\lstdefinestyle{mystyle}{
frame=single,
    language=Java,
    aboveskip=3mm,
    belowskip=3mm,
    showstringspaces=false,
    escapeinside={(*@}{@*)},
    columns=flexible,
   basicstyle={\large \ttfamily},
    numbers=none,
    numberstyle=\large\color{black},
    keywordstyle=\color{blue},
    commentstyle=\color{dkgreen},
    stringstyle=\color{mauve},
    otherkeywords = {@BeforeEach,@Test, MIN_VALUE, MAX_VALUE},
    morekeywords = [2]{decreases, invariant , print}, morekeywords = [3]{assertTrue, assertFalse,assertEquals,assertSame, assertAll}, morekeywords = [4]{@BeforeEach}, breaklines=true,
    breakatwhitespace=true,
    tabsize=4
}

\newcommand{\revise}[1]{
   \textcolor{black}{#1}
}

\newcommand{\CodeGreenHilight}{\makebox[0pt][l]{\color{green!20}\rule[-0.45em]{\linewidth}{1.3em}}}

\newcommand{\CodeRedHilight}{\makebox[0pt][l]{\color{red!20}\rule[-0.45em]{0.85\linewidth}{1.3em}}}

\def\tsc#1{\csdef{#1}{\textsc{\lowercase{#1}}\xspace}}
\tsc{WGM}
\tsc{QE}

\begin{document}
\let\WriteBookmarks\relax
\def\floatpagepagefraction{1}
\def\textpagefraction{.001}

\shorttitle{}    

\shortauthors{M.R.H Misu et al.}  

\title [mode = title]{Test Smell: A Parasitic Energy Consumer in Software Testing}  

\tnotemark[1]

\author[1]{Md Rakib Hossain Misu}[
        orcid=0000-0002-7931-6782
        ]
\cormark[1]
\fnmark[1]
\ead{mdrh@uci.edu}

\author[1]{Jiawei Li}[
orcid=0000-0002-4434-4812
]
\fnmark[2]
\ead{jiawl28@uci.edu}
\author[1]{Adithya Bhattiprolu}\fnmark[3]
\ead{abhattip@uci.edu}
\author[1]{Yang Liu}\fnmark[4]
\ead{yangl73@uci.edu}

\author[2]{Eduardo Almeida}[
    orcid=0000-0002-9312-6715
]
\fnmark[5]
\ead{esa@rise.com.br}

\author[1]{Iftekhar Ahmed}[
    orcid=0000-0001-8221-5352
]
\fnmark[6]
\ead{iftekha@uci.edu}

\affiliation[1]{organization={University of California, Irvine},
            addressline={}, 
            city={Irvine},
state={CA},
            country={USA}}
\affiliation[2]{organization={Federal University of Bahia},
            addressline={}, 
            city={Salvador},
state={Bahia},
            country={Brazil}}

\cortext[1]{Corresponding author}

\fntext[1]{}

\begin{abstract}
\textit{\textbf{Context:}} Traditionally, energy efficiency research has focused on reducing energy consumption at the hardware level and, more recently, in the design and coding phases of the software development life cycle. However, software testing's impact on energy consumption did not receive attention from the research community. Specifically, how test code design quality and test smell (e.g., sub-optimal design and bad practices in test code) impact energy consumption has not been investigated yet.\\
\textit{\textbf{Objective:}} This study aims to examine open-source software projects to analyze the association between test smell and its effects on energy consumption in software testing.\\
\textit{\textbf{Methods:}} We conducted a mixed-method empirical analysis from two perspectives; software (\revise{data mining in 12 Apache projects}) and developers' views (\revise{a survey of 62 software practitioners)}.\\ 
\textit{\textbf{Results:}} Our findings show that: 1) test smell is associated with energy consumption in software testing. Specifically, the smelly part of a test case consumes more energy compared to the non-smelly part. 2) certain test smells are more energy-hungry than others, 3) refactored test cases tend to consume less energy than their smelly counterparts, and 4) most developers \revise{(45\% of the survey respondents)} lack knowledge about test smells' impact on energy consumption.\\
\textit{\textbf{Conclusion:}} 
Based on the results, we emphasize raising developers awareness regarding the impact of test smells on energy consumption. Additionally we present several observations that can direct future research and developments.
\end{abstract}


\maketitle
\section{Introduction}
\label{sec:introduction}
With the popularization of personal computers and mobile devices, millions---if not billions---of software applications in those devices have been governing countless aspects of our lives in recent years. These software applications consume energy when being used and developed. Based on a report by Huawei Technologies~\cite{andrae2017total}, information and communication technology consumed around 9\% of the total global energy in 2018 where software plays a major part. In addition, software-related products and services have already become one of the major contributors (2.3\%) of global Green House Gases~\cite{SMART} and growing at a much faster rate than initially predicted \cite{gesi,DBLP:journals/csur/GeorgiouRS19}. Therefore, it becomes critical to control energy consumption and reduce Green House Gas production due to software usage and development to ensure a better and more sustainable future~\cite{Unitednations}.
\par
The primary focus of research pertaining to energy-efficient solutions has been on the optimization of the hardware with the aim of making it more energy-efficient for executing software. \cite{DBLP:conf/ics/BircherJ08,DBLP:conf/iccad/IyerM02,DBLP:conf/eurosys/MerkelB06, DBLP:conf/isca/RanganWB09}. More recently, energy consumption incurred by software executions and software runtime performance \cite{DBLP:conf/asplos/RibicY14, DBLP:conf/jvm/VijaykrishnanKKTSI01}, especially in mobile \cite{DBLP:conf/greens/BehrouzSGMA15,DBLP:conf/icsm/KwonT13} and embedded software \cite{DBLP:journals/tvlsi/TiwariMW94}, received more attention from the research community as energy efficiency is crucial for mobile applications and embedded systems. Another thread of research investigated various aspects of the Software Development Life Cycle (SDLC) and their association with energy efficiency \cite{DBLP:conf/msr/PintoCL14, DBLP:journals/csur/GeorgiouRS19}. For instance, in the design and coding phase, following a software architecture, applying a design pattern \cite {DBLP:conf/icse/ChowdhuryHKSMK19,DBLP:journals/smr/FeitosaAAAN17,DBLP:journals/jss/VentersCBPCCNBC18}, adopting a specific programming language and framework all have impact on energy consumption~\cite{DBLP:conf/sblp/0001PRRS17,DBLP:conf/sle/Pereira0RRCFS17,DBLP:conf/greens/PereiraCSCF16,CALERO2021100603}. 
\par
Prior work however is limited to only some phases of the SDLC and provides a fragmented view of the possible energy-efficient techniques associated with the SDLC. For example, none of the existing work has looked into the effect of testing on energy consumption even though millions of lines of test code that test various aspects of software \cite{DBLP:journals/tse/PetrovicIFJ22} are being executed daily in developers' IDEs or in the Continues Integration (CI) pipeline, which we believe could consume a considerable amount of energy on a daily basis.
\par
In this study, we posit that a sub-optimal test case design ( a.k.a, a test smell~\cite{van2001refactoring,DBLP:conf/icst/GreilerDS13,DBLP:conf/cascon/PerumaANM0P19}) has the potential to incur more unnecessary energy overhead than an optimal design given the same production code under testing. We aim to investigate the impact of test smells on energy consumption during software testing.

\par
Figure \ref{fig:energy-smell} demonstrates a motivating example that shows how the presence of test smells can significantly increase the energy consumption of running the test suite. The example code snippet represents a General Fixtures (\texttt{GF}) test smell~\cite{DBLP:journals/ese/BavotaQOLB15} that occurs when a test class contains a \texttt{setUp()} method that may not be directly relevant to the executed test case, that is, the test case \texttt{testRangeOfChats()} has never utilized any field variables initialized in the \texttt{setUp()} method. However, the \texttt{setUp()} is invoked before every execution of this test case. Consequently, the presence of this test smell can lead to additional computations and memory usage because of unnecessary setup and teardown operations. While the direct impact may not be significant for individual test runs, the cumulative effect can substantially affect energy consumption in large-scale software projects with extensive test suites and frequent test executions.

\begin{figure*}[t]
\centering
\begin{tcolorbox}
[   left=2pt,
    width=0.75\textwidth, 
    colframe=black, 
    colback=anti-flashwhite!0, 
    boxsep=2mm, 
    arc=2mm,
    title={General Fixture (GF) Test Smell},
    skin=bicolor,
    bottom=0mm,
    height=10cm,
    boxrule=0.5pt,
    fontupper=\large
]
\vskip-5mm
\begin{lstlisting}[style = mystyle, frame=none, label=lst:example, basicstyle={\small \ttfamily}]
36: @SuppressWarnings("boxing")
36: public class IntegerRangeTest extends AbstractLangTest {
...     .............
47 :     private IntegerRange range1, range2, range3, rangeFull;
....     
....    /**Initializing field variables before each test method**/
55 :     @BeforeEach
56 :     public void setUp() {
57 :         range1 = of(10, 20);
58 :         range2 = of(10, 20);
59 :         range3 = of(-2, -1);
60 :         rangeFull=of(Integer.MIN_VALUE,Integer.MAX_VALUE);
61 :     }
....     .............
....     .............
....     /** Field variables were never utilized in test method**/
386:     @Test
387:     public void testRangeOfChars() {
389:         final IntegerRange chars = of('a', 'z');
390:         assertTrue(chars.contains((int) 'b'));
391:         assertFalse(chars.contains((int) 'B'));
392:     }
.... }
\end{lstlisting}
\end{tcolorbox}

\captionsetup{justification=centering, labelsep=colon, name=Figure}
\caption{An Example of Smelly Test from \texttt{org.apache.commons.lang3.IntegerRangeTest} class in Apache Commons-Lang.}
\label{fig:energy-smell}
\vspace{-5mm}
\end{figure*}

\par
A plethora of studies have investigated test smells' impacts on software maintainability, comprehension, and defect proneness\cite{DBLP:series/springer/MoonenDZB08,DBLP:conf/icsm/BavotaQOLB12}. Researchers also investigated automated test smell detection \cite{DBLP:conf/sbes/VirginioMSSCCM20,DBLP:conf/kbse/WangGSLB021, DBLP:conf/kbse/TaniguchiMK21}, and refactoring \cite{DBLP:journals/tse/SoaresRGAS23, DBLP:conf/iwpc/LambiaseCPLP20,DBLP:conf/msr/NagyA22}. However, to the best of our knowledge, no study has investigated the association between test smell and software energy consumption incurred during software testing with the same production code under test. Given the widespread prevalence of test smells \cite{DBLP:conf/icsm/BavotaQOLB12}, and developers' unawareness regarding the relation between test smell and energy consumption as indicated by some of the respondents in our survey \textit{"I have no idea how these things are related to each other"[S-6] \footnote{Here [S-6] refers to our survey respondent's anonymous Id.} or  \textit{"They seem like two unrelated concepts."[S-17]}}, it is of utmost importance that developers and researchers are aware of the relationship between test smell and energy consumption.
\par
In this paper, we aim to contribute to the literature by providing a comprehensive study on the impact of test smells on software energy consumption with the production code under test unchanged. We design our study from two perspectives, \textit{software} and \textit{developer}. We first investigate whether the smelly test code in a test case consumes more energy than its clean part. If so, does a test case with more test smell instances consume more energy than the ones with less smell instances?
\par
Besides, we also conduct a case study where we manually removed the test smells to create test cases without smells. We analyze the energy consumption difference between clean and smelly test cases that test the same production code to explore the impact of test smells on energy consumption during software testing. To further understand different test smell types' impacts on energy consumption, we perform a correlation analysis between the number of instances of each test smell type and an estimate of energy consumption per smell instance. The goal is to see which test smell types are the most energy-hungry.
\par
In addition, we also aim to understand the software developers' awareness of test smell, its impact on energy consumption and the possible reasons that developers introduce test smells that cause unnecessary energy overhead.
Specifically, we answer the following research questions:
\vskip +2mm

\textbf{RQ1~[Test Smells vs. Energy]:} \textit{How do smelly tests, in general, impact energy consumption?}
\vskip +2mm
\textbf{RQ2~[Test Smell Types vs. Energy]:} \textit{How does each test smell type impact energy consumption?}
\vskip +2mm
\textbf{RQ3~[Developers' Awareness]:} \textit{Are developers aware of the impact of test smells on energy consumption?}
\vskip +2mm
\textbf{RQ4~[Provenance of Test Smell]:} \textit{What are the underlying reasons for developers to introduce test smells that could cause unnecessary energy consumption?}

\vskip +3mm
Overall this paper makes the following contributions:
\begin{enumerate}
    \item [\faUnlock*] We conduct the first study to investigate the affect of test smell on energy consumption during software testing.
    \item [\faUnlock*] We recognize that the presence of certain smells in the test code leads to additional energy consumption in software testing.
    \item [\faUnlock*] We present the findings of a survey of test smell's impact on energy consumption with 62 software practitioners. The survey reflects developers' perceptions and actions about test smells.
    \item [\faUnlock*] We find that all developer groups (i.g, core, non-core, and bot) almost equally contribute to introducing energy-hungry test smells.
\end{enumerate}

\par
The paper is structured as follows: we describe the prior research on test smell's impact and software energy efficiency in Section \ref{sec:relatedworks}, followed by our approach of detecting test smells in software repositories, profiling energy consumption, and surveying developers for understanding developers' views in Section \ref{sec:method}. In Section \ref{sec:results}, we present our analysis and finding, and Section \ref{sec:discussion} provides the results' implications for researchers and software developers.

 \section{Related Work}
\label{sec:relatedworks}
\subsection{The Impact of Test Smells}
\revise{Test smells refer to symptoms of sub-optimal design choices or bad programming practices in software test code~\cite{van2001refactoring}. Researchers have defined various types of test smells that occur in large software projects, such as Assertion Roulette (\texttt{AR}), Lazy Test (\texttt{LT}), and Mystery Guest (\texttt{MG}) etc~\cite{van2001refactoring,DBLP:conf/icst/GreilerDS13,DBLP:conf/cascon/PerumaANM0P19}. These smells have been proven to have a negative effect on test quality~\cite{DBLP:conf/icst/HassanR22}. The impacts of test smell on software readability, understandability, maintainability, and performance have also been widely studied in literature~\cite{DBLP:conf/kbse/TufanoPBPOLP16,DBLP:conf/sbes/VirginioSMSCM19}. For example, Bavota et al.~\cite{DBLP:journals/ese/BavotaQOLB15} conducted experiments to investigate the impact of test smells on program comprehension. Their results showed that test smells have a negative impact on both the comprehensibility and maintainability of the test code. Spadini et al.~\cite{DBLP:conf/icsm/SpadiniPZBB18} found that smelly test cases are more change and defect-prone, and they could cause the tested production code to be more defect-prone. However, no existing research has investigated the impacts of test smells on energy consumption in software testing. In this study, we contributed to the literature by studying the affects of test smells on energy consumption in software testing.}

\subsection{Software Energy Efficiency}
Building software that is more energy-efficient has become an integral concern in improving sustainability. In the past, researchers have been trying to identify the factors that might lead to energy inefficiencies in software ~\cite{DBLP:conf/msr/VasquezBBOPP14,DBLP:conf/sigsoft/SongZH19,DBLP:conf/msr/OliveiraOCF019,DBLP:journals/tosem/LiCLWG23,DBLP:conf/sigsoft/JabbarvandM17,DBLP:conf/issta/BehrouzSBM16,DBLP:conf/esem/SahinPC14,DBLP:conf/icsm/SahinTMPC14}. Liu et al.~\cite{DBLP:journals/tse/LiuXCL14} found that wakelock deactivation and missing sensors as two of the main causes of energy inefficiencies in Android applications. Baberhee et al.~\cite{DBLP:conf/icse/BanerjeeR16} proposed several guidelines to refactor Android apps affected by energy-oblivious design practices, such as balancing the quality of service and functionality and restricting resource leaks. Bruce et al.~\cite{DBLP:conf/gecco/BrucePH15} employed search-based software engineering techniques to automatically identify more energy-efficient versions of the MiniSAT Boolean satisfiability solver. Manotas et al.~\cite{DBLP:conf/icse/ManotasPC14} built a framework that improves the energy efficiency of Java software by automatically using the most energy-efficient library implementations. Such energy-aware implementations made in practice have also been explored and analyzed by Moura et al.~\cite{DBLP:conf/msr/MouraPEC15}, which suggests that developers mostly utilize low-level energy management approaches such as idleness.
\par
More recently, Song et al.~\cite{DBLP:conf/sigsoft/SongZH19} defined four
anti-patterns of service usage inefficiency in Android applications, including premature create, late destroy, premature destroy, and service leak, which could lead to high energy consumption. Song et al.~\cite{song2021empirical} found that the high average energy consumption is due to some methods that are frequently invoked by test cases (i.e., energy hotspots). In addition, Li et al.~\cite{DBLP:journals/tosem/LiCLWG23} explored various root causes of energy issues in mobile apps, such as unnecessary workload and wasted background processing. However, none of the existing works have investigated the anti-patterns/bad practices in software testing that could contribute to energy overhead, and we aim to fill that gap. \begin{figure*}[t]
\centering
\includegraphics[width=0.90\linewidth]{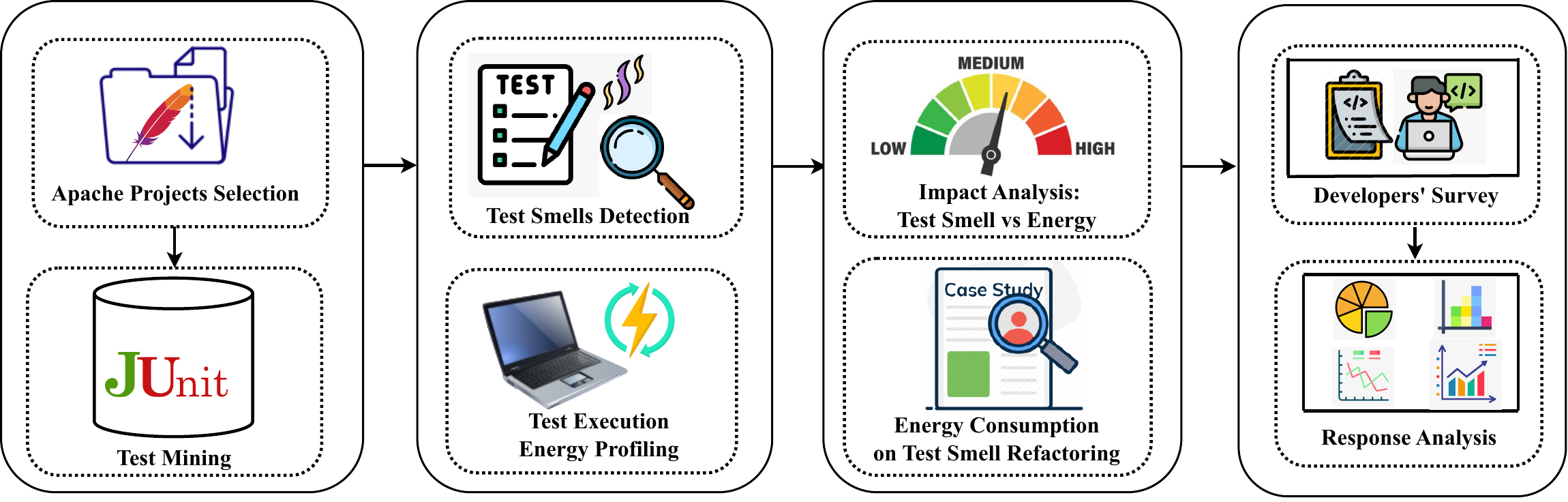}
\captionsetup{justification=centering, labelsep=colon, name=Figure}
\centering
\caption{Overview of The Study Design with Experimental Phases}
\label{fig:methodology}
\end{figure*}

\section{Methodology}
\label{sec:method}
This study aims to understand the impact of test smells on energy consumption during software testing.
Figure \ref{fig:methodology} demonstrates a high-level overview of our study. In the following sections, we provide a comprehensive description of each phase of our experiment.

\subsection{Subject Systems}
To conduct our study, we looked for software projects that satisfy our experimental attributes and criteria. We started our experiment with a collection of open-source Java software projects from the Apache Software Foundation
~\cite{apache-software-foundation}. We decided to select these projects for three reasons. First, we chose the Apache projects as these projects are well-maintained and supported by a large developer community. Besides, prior research works have been conducted on these Apache projects ~\cite{ DBLP:conf/msr/KabinnaBSH16}. Second, we selected projects written in Java since it is considered one of the most widely-used programming languages~\cite{TIOBE}. Third, in the literature, most of the test smell detection tools are available for Java compared to other languages \cite{10.1145/3463274.3463335}. For our analysis, we needed to build the projects and execute the test cases. To avoid complications related to building a project, we selected projects using Maven~\cite{maven} as its build system. Next, we selected projects that use JUnit~\cite{massol2004junit} as the unit testing framework, since it is one of the most widely-used unit testing frameworks for Java. We identified 25 projects that met all these criteria.
\par
We decided to perform our analysis at the test case granularity to analyze test code quality and measure energy consumption. According to
JUnit documentation, a method in the test code with \texttt{@Test} annotation represents a test case, and it can be executed inside the
complete test suit or explicitly invoked with its Fully Qualified Name (\texttt{FQN}) from the command line. 
To locate a test case and its Lines Of Code (LOC), we employed the static code analysis technique utilizing the Eclipse JDT Core Abstract Syntax Tree (AST) traversal tool~\cite{eclipse-jdt}. For each project, we ran each the test cases individually with the help of Maven and JUnit. During the execution, we discarded a project if one of its test cases failed to execute independently.
\par
We recognized a subset of 12 projects where all test cases can independently execute from the command line. Finally, we ended up with these 12 Apache Java projects as our experimental projects containing 13,103 test cases in total.

 \subsection{Test Smell Detection}
\label{sec:test-smell-detection}
In the literature, researchers have proposed various test smell detectors where we found 18 tools that can detect test smells in JUnit-based Java projects~\cite{10.1145/3463274.3463335}. We observed that among these tools, tsDetect~\cite{DBLP:conf/sigsoft/PerumaANM0P20} can detect the presence of 19 types of test smells, achieving the highest precision and recall \cite{DBLP:conf/sigsoft/PerumaANM0P20, 10.1145/3463274.3463335}. However, in the test smell detection report, tsDetect does not provide the locations of the smells in the test code.
\par
In our study, we are interested in analyzing both the smelly and non-smelly parts of the test code. Therefore, we searched for an extension of tsDetect that can also identify test smell locations in the test code. We encountered JNose~\cite{DBLP:conf/sbes/VirginioMSSCCM20}, which has reused the same test smell detection rules employed in tsDetect and can detect 21 types of test smells, including the 19 types of tsDetect detected smells. In addition, JNose provides the locations (i.e., line numbers) and test case names where a test smell is identified in the test code and the number of each type of test smell identified in a test class.
\par 
To assess the correctness of JNose in terms of precision and recall, we utilized a benchmark of 65 JUnit test files containing instances from various smell types. This benchmark has been created and employed in an earlier qualitative study to evaluate tsDetect~\cite{DBLP:conf/sigsoft/PerumaANM0P20}. We executed JNose on that benchmark dataset and compared its test smell detection results with tsDetect. We observed that both JNose and tsDetect got the same overall precision score ranging from 85\% to 100\% and a recall score from 90\% to 100\% with an average F-score of 95\%. Besides, we also found that JNose has successfully been adopted by researchers in recent studies \cite{DBLP:journals/ese/SharmaGKGS23,martins2023diffusion}. This inspired us to utilize JNose in our experiment. A summary of JNose-detected test smells is demonstrated in Table \ref{tab:test-smell-types}.
\par
To detect and locate test smells in a given repository, JNose first parses the source code into Abstract Syntax Tree (AST) and then traverses the AST applying detection rules for identifying test smells \cite{DBLP:conf/sbes/VirginioMSSCCM20}. Once the detection is completed, JNose generates a report containing information, for instance, test class, number of different types of smells, smelly test cases, and the source code line numbers where the test smell appears. Our experiment requires detecting the presence of different test smells in each test case. To do so, we executed JNose on the subject systems. Utilizing these reports, we extracted the instance of various test smells found in a test case and counted the total number of smell instances, Smell Count ($SC$). In total, we detected 56,908 test smell instances in 12 projects, and the frequency of each type of smell is shown in Table \ref{tab:test-smell-types}. We recognized that some types of test smells occur in a single line in the test case such as Assertion Roulette (\texttt{AR}), Redundant Assertion (\texttt{RA}) etc, whereas Lazy Test (\texttt{LT}), Eager Test (\texttt{ET}), Conditional Test Logic (\texttt{CTL}) some other smells occur in multiple lines. To quantify the portion of smelly and clean (i.g, non-smelly) test code, we parsed the location where these smells were identified and calculated the smelly line of code ($LOC_{(smell)}$) and clean line of code ($LOC_{(clean)}$) of a test case. At the end of this phase, we created a tuple of the test case, $LOC$, $LOC_{(smell)}$, $LOC_{(clean)}$, $SC$, and the count of each type of test smell, such as:
\par
$testcase \rightarrow$\\
$\{LOC, LOC_{(smell)}, LOC_{(clean)}, SC, s_{1}, s_{2}, s_{3},...s_{21}\}$.

\begin{table*}[t]
\centering
\caption{JNose detected test smells and their frequency in 12 Apache projects} 
\label{tab:test-smell-types}
\begin{tabular}{|l|l|l|p{6.8cm}|c|}
\hline
\textbf{ID} & \textbf{Test Smell Name}    & \textbf{Abbreviation} & \multicolumn{1}{c|}{\textbf{Definition}}                                                                                                                  & \textbf{Frequency} \\ \hline
$s_{1}$ & Assertion Roulette          & \texttt{AR }   & Appears when a test case contains several assertions without explanation messages. & 42.32\%              \\ \hline
$s_{2}$ & Lazy Test                   & \texttt{LT}                    & Appears when multiple test cases check the same method of a production object  &   26.50\%  \\ \hline
$s_{3}$ & Magic Number Test           & \texttt{MNT}   & Appears when a test case contains undocumented numerical values  &    11.36\%           \\ \hline
$s_{4}$ & Eager Test                  & \texttt{ET}    & Occurs when a test case invoking multiple methods of the product object to be tested &    5.72\%           \\ \hline
$s_{5}$ & Duplicate Assert            & \texttt{DA}                    & Appears when a test case has the identical assertion multiple times.&   2.95\%            \\ \hline
$s_{6}$ & Sensitive Equality          & \texttt{SE}                    & Appears when a test case has an assertion that checks equality with toString method  &  2.34\%   \\ \hline
$s_{7}$ & Conditional Test Logic & \texttt{CTL}  & Occurs when a test case contains a conditional statement like (i.e., if/else ) as a prerequisite to executing other test statements & 2.20\%              \\ \hline
$s_{8}$ & Unknown Test                & \texttt{UT}    & Appears when a test case has no assertions statement or non-descriptive name &     2.18\%          \\ \hline
$s_{9}$ & Exception Handling & \texttt{EH} & Found when a test case includes custom exception handling instead of utilizing JUnit’s exception handling features &   2.08\% \\ \hline
$s_{10}$ & General Fixture & \texttt{GF}  & Emerges when in a test class, the \texttt{setUp()} fixture creates many objects, and the test case only uses a subset of those objects. &   0.36\%   \\ \hline
$s_{11}$ & Redundant Assertion         & \texttt{RA}    & Occurs if a test case contains an assertion statement that is permanently true or false & 0.29\%           \\ \hline
$s_{12}$ & Ignored Test                & \texttt{IgT}                   & Occurs when a test case has an ignore annotation that prevents the test case's execution. &  0.24\%   \\ \hline
$s_{13}$ & Constructor Initialization  & \texttt{CI}                    & A test class contains a constructor declaration &   0.23\% \\ \hline
$s_{14}$ & Resource Optimism           & \texttt{RO}                    & Emerges when a test case assumes the existence of external resources. &   0.19\%            \\ \hline
$s_{15}$ & Mystery Guest               & \texttt{MG}    & Exists if a test case accesses external resources such as a database, directory, or file that contains test data. & 0.17\% \\ \hline
$s_{16}$ & Verbose Test                & \texttt{VT}                    & Occurs if a test case runs additional staff and becomes complex &   0.80\% \\ \hline
$s_{17}$ & Sleepy Test                 & \texttt{ST}    & Occurs when a test case calls an explicit wait like Thread.sleep(). &    0.02\%           \\ \hline
$s_{18}$ & Empty Test      & \texttt{EmT}  & Found when a test case  has an empty body or does not have executable any statements &    0.02\%           \\ \hline
$s_{19}$ & Redundant Print
           & \texttt{RP}                    & Occurs when a test case contains print statements.      &   0.01\%  \\ \hline
$s_{20}$ & Dependent Test              & \texttt{DepT}  & Arises when a test case only runs on the successful execution of other test cases. &    0.00\%           \\ \hline
$s_{21}$ & Default Test                & \texttt{DT}                    & Default or scaffold test cases created by IDEs or frameworks. &   0.00\%            \\ \hline
\end{tabular}
\end{table*}

 \subsection{Energy Measurement}
\label{sec:energy-measurement}
In this stage, we executed each test case with a energy profiler that monitors the energy consumption rate during its execution. 

\subsubsection{\textbf{Experimental Environment Setup}} To create an experimental environment and conduct our analysis, we used five MacBook Airs (13-inch, Mid-2012) laptops. All these laptops have the same hardware configuration containing a 1.8GHz dual-core Intel® Core™ i5 processor, 8GB RAM, and 256 GB SSD running macOS Catalina 10.15.7. To support the build configuration and test case execution of projects, we used Maven 3.8.6 build system.

\subsubsection{\textbf{Energy Profiler}} 
To monitor the actual energy usage of a device at any given time, the most precise method is to employ power meters which are hardware tools connecting the device's power sources. Although power meters offer high accuracy, these tools are challenging to configure and require customized changes to the device under-monitored. Therefore, power meters are more suitable for mobile devices where specific hardware can easily be accommodated by connecting external chips or sensors. Besides, to measure energy consumption for a specific software running on a device, the power meter needs to be synchronized with software execution time avoiding noises from other background services and external factors. To monitor energy consumption in software execution, the suggested approach~\cite{cruz2021tools} is to employ an energy profiler that computes power cost based on an estimation model of the power cost of various hardware modules. Depending on which modules are active during the software execution, the profiler estimates a particular energy cost.
\par
In our experiment, we used Intel PowerLog profiler to measure the energy consumption~\cite{intel,DBLP:conf/vl/Carcao14}. It is a command-line tool provided by the Intel Power Gadget toolkit, a power usage monitoring tool. Intel PowerLog precisely estimates power usage from a software level without any hardware instrumentation \cite{intel}. It is supported on macOS to monitor and assess real-time processor package power information using the energy counters in the Intel® Core™ processors\cite{intel}. The primary motivation for adopting Intel PowerLog is that it provides a convenient technique to measure processor power usage while executing a specific command in the command line and storing energy profiling into a log file. The log file contains values representing energy, power, and the total time duration in a sequence of times lap, including the total energy consumption $E(Joules)$, required power $P(Watt)$, and entire time duration $T(sec)$, respectively.
\par
Since Intel PowerLog measures energy metrics while executing a specific user command at the process level, it minimizes the effects of other processes. Therefore, its estimations do not include energy consumed by main memory or I/O process~\cite{DBLP:conf/gecco/BrucePH15}
Moreover, Intel PowerLog has been used in prior research and suggested by the researchers~\cite{DBLP:conf/gecco/BrucePH15, DBLP:conf/vl/Carcao14,DBLP:journals/corr/abs-2401-06482, cruz2021tools,cruzbekker2023energyunits}, increasing the confidence in using this for our analysis. We installed the latest Intel PowerLog version: \texttt{3.7.0} on all these MacBooks.

\subsubsection{\textbf{Test Case Execution}}
We ran each test case with the \texttt{mvn test -Dtest="$<$test case FQN$>$"}
command while monitoring energy usage using the Intel PowerLog. To minimize external interference, we configured the laptops to Zen Mode \cite{measuring-energy} before executing the test cases which prevents these laptops from interacting with external networks, and devices. We maintained the following configurations on each laptop to keep it in Zen Mode during the execution of the test cases.

\begin{enumerate}
    \item[\faLaptopCode] We fully charged the laptop to 100\%~(\faBatteryFull). However, to provide equal battery capacity for test case runs, we kept the laptop plugged~(\faPlug) in throughout the experiment.
    \item[\faLaptopCode] All active applications were quit and unnecessary background services were killed except the terminal (\faTerminal). The auto-dim of an inactive screen was turned off. The screen saver was set up to appear for one hour, and the sleep time was set \textit{never} to prevent the laptop from falling into sleep mode. Microphone (\faMicrophoneSlash) and Speaker (\faVolumeMute) were also turned off.
    \item[\faLaptopCode] Automatic adjusting of brightness was turned off. Brightness~(\faBahai) was also lowered to 50\%, and the keyboard lighting, automatic logged out, notifications, AirDrop, Bluetooth~(\faBluetoothB), and WiFi~(\faWifi) were turned off.
\end{enumerate}
\par
In order to obtain accurate energy measurements, it was necessary to execute a test case multiple times with the Intel PowerLog while the assigned laptop was in Zen Mode. We conducted a preliminary experiment to determine the ideal number of test case runs for reliable energy measurements. Initially, we employed a stratified random sampling, aiming for a 90\% confidence interval and a 10\% margin of error that led us to select 68 test cases out of 13,103 test cases. Subsequently, we designed a script that automates the execution of a test case with Intel PowerLog for a predetermined number of iterations, incorporating a 30-second cool-down interval after each run. The cool-down period prevents both tail energy consumption from the previous measurement and the collateral tasks of the last execution from affecting the subsequent measurement.
\par
Next, we executed that script to run each sampled test case 5, 15, and 25 times. From the resulting log files, we individually extracted the median values of energy $E (Joules)$, power $P (Watts)$, and time $T (seconds)$. We generated a plot illustrating the relationship between the number of runs and the median energy values. Our analysis revealed that the median values exhibited insignificant variances when the test cases were executed 5, 15, and 25 times, indicating overall consistency. Based on these experimental findings, we determined that the number of runs had no significant impact on the median energy consumption. Consequently, we proceeded with five runs of each test case.
\par
In total, we executed 13,103 test cases five times, each with 30 second cool-down period. The whole experiment took approximately 874 hours (37 days) of execution time on each laptop. We then extracted five generated energy log files for each test case and calculated the median value of energy, power, and time. Thus, for a project, we created a list of test cases with their corresponding energy $E(Joules)$, power $P(Watt)$, and time $T(sec)$ values in the form of tuples such as: $testcase \rightarrow \{E, P, T\}$.
 \subsection{Impact Analysis: Test Smell vs. Energy Consumption}
During test smell detection and energy measurement, we generated two types of tuples for the test cases. We then joined them and created a list of test cases with their smell and energy-associated values. This list of tuples helps us analyze test smell's impact on energy consumption.\\
$testcase \rightarrow$ \\
$\{LOC, LOC_{(smell)}, LOC_{(clean)}, SC, E, P, T, s_{1},s_{2},...s_{21}\}$.
\\

\subsubsection{\textbf{Group Analysis}} 
\label{sec:group-analysis}
We first investigated smelly tests' impact on energy consumption. Our goal in this step is to see if the test smells are associated with increased energy consumption. To do so, for each test case, we calculated the energy $E_{LOC_{(smell)}}(Joules)$ required to execute a smelly line of code ($LOC_{(smell)}$)  and the energy $E_{LOC_{(clean)}}(Joules)$ for clean line of code ($LOC_{(clean)}$). According to Equations \ref{eq:1} and \ref{eq:2}, we computed these energy values.
We believe these two values could serve as estimates for the energy consumption of smelly and clean parts of the test code in a test case. Next, we categorized all the test cases into multiple groups based on their total smell count ($SC$). The goal is to analyze if more test smell instances increase the energy overhead incurred by test smells. We created these groups with the interval of various smell counts, such as 5, 10, and 25 to mitigate the bias brought upon by selected group size. Then for each group, we calculated mean and median energy consumption of $E_{LOC_{(smell)}}$ to analyze the trend of energy consumption. We also conducted Welch's t-test \cite{ruxton2006unequal,cohen2014applied} and Cohen's D~\cite{diener2010cohen} to measure the statistical significance and effect size of the differences in energy consumption between different groups. 

\begin{equation} 
\label{eq:1}
E_{LOC_{(smell)}} = E*\frac{LOC_{(smell)}}{LOC_{(test)}} (Joules)
\end{equation}

\begin{equation} 
\label{eq:2}
E_{LOC_{(clean)}} = E* \frac{LOC_{(clean)}}{LOC_{(test)}} (Joules)
\end{equation}

\begin{equation} 
\label{eq:3}
E_{(N)} = \frac{E}{SC} (Joules)
\end{equation}
\\

\subsubsection{\textbf{Correlation Analysis}} 
\label{sec:correlation-analysis}
To further understand how individual test smell type impacts energy consumption, we investigate the relationship between energy consumption and each test smell type. However, \revise{we do not} have a one-to-one mapping between smell type vs. energy consumption to establish the relationship. We illustrate it by provide an example: Supposes a test case contains 12 test smells (e.g., $SC=12$) in 3 different types, for example, $S_{1}=7$, $S_{4}=3$, and $S_{15}=2$. The energy consumption ($E$) reflects the complete execution of that test case with the presence of all those 12 test smells. Hence, it is not possible to differentiate which test smell affects what portion of the energy consumption. Therefore, establishing a relation between ($E$ vs. $S_{1}$),($E$ vs. $S_{4}$), and ($E$ vs. $S_{15}$) would not be accurate.
To normalize the energy consumption for each smell, we measured the energy consumption per test smell using $E_{(N)}$ according to Equation \ref{eq:3}, where SC is the total smell count. For a test case, the normalized value $E_{(N)}$ represents the energy consumption per test smell that relates to its effect on energy consumption. With this one-to-one mapping, we conducted a Kendall's Tau ($\tau_{b}$) correlation analysis \cite{cohen2014applied} between energy consumption per test smell and the number of test smell instances for each test smell type. We used Kendall's Tau since it has smaller gross error sensitivity and smaller asymptotic variance compared to Spearman correlation~\cite{sen1968estimates}.

 \subsection{Impact Analysis: Case Study of Test Smell Refactoring}
\label{casestudy}
To gain a comprehensive understanding of the impact of test smells on energy consumption, we conducted a case study. We began by selecting the
largest project from our pool of subject systems. Next, we sampled test cases and manually refactored them to eliminate any existing test smells,
creating test case pairs (i.e., with and without test smells) that test the same production code functionality. We executed these refactored test
cases using an energy profiler to measure the energy consumption rate. The following subsections provide details of each step undertaken in the
case study.

\subsubsection{\textbf{Subject Selection}}
In order to refactor the test cases of a project, it is essential to possess a reasonable comprehension of the project's codebase and its test suite. Nevertheless, comprehending the codebase of all 12 projects from our subject systems and refactoring 21 distinct types of smells in thousands of test cases present a challenging and demanding task. Therefore, we have opted to refactor a random sample of smelly test cases in a single project, which makes the codebase understanding and manual test smell refactoring manageable. For this purpose, we have selected the Apache Commons Lang project due to having the largest number of test cases (4,067), lines of code (89K LOC), and smelly test cases (2,032) among all our subject systems.

\subsubsection{\textbf{Sampling Test Cases}}
Existing literature has demonstrated that certain test smells tend to co-occur together. Refactoring test smells like Lazy Test (\texttt{LT}), Eager Test (\texttt{ET}), and Conditional Test Logic (\texttt{CTL}) often involves modifying multiple test cases or introducing new ones. Consequently, developers may perform partial or complete refactoring. In partial refactoring, developers address some test smells in a test case, reducing the overall number of smells. On the other hand, complete refactoring involves removing all types of smells, resulting in a clean test case. To analyze the impact of energy consumption in both of these situations, we created two types of sampled test cases. These test cases represent the scenarios of partial refactoring, where only some test smells are addressed, and complete refactoring, where all test smells are removed, allowing us to study the energy consumption variations comprehensively.
\par
To analyze the energy impact of partial refactoring, we created a stratified sample of test cases that contain different types of smells. We first determined the size of the random sample by utilizing a 90\% confidence interval and a 10\% margin of error, giving us a sample size of 66. In the Apache Commons Lang project, there were a total of 2,032 smelly test cases. We proceeded to select a stratified random sample of 66 test cases from this pool of 2,032 smelly test cases. These 66 test cases contain various types and numbers of test smells. We tagged them as ``\texttt{Smelly-66}''.
\par
To observe the energy consumption impact of complete refactoring, we filtered out test cases that contained test smells highly associated with energy consumption. Our analysis in Section \ref{sec:rq2-result} revealed that certain test smells, namely Assertion Roulette (\texttt{AR}), Lazy Test (\texttt{LT}), and Eager Test (\texttt{ET}), have high correlations with energy consumption, with \texttt{AR} having the highest correlation. However, refactoring \texttt{ET} and \texttt{LT} may require changes in multiple test cases or the introduction of new ones which goes beyond the scope of our study. Therefore, we focused on refactoring test cases that solely contained Assertion Roulette (\texttt{AR}) smells. After filtering, we identified 79 test cases out of the initial 2,032 that only contained \texttt{AR} smells. We tagged these test cases as ``\texttt{Smelly-AR-79}'' for further analysis.

\begin{figure}[t]
\includegraphics[width=\linewidth]{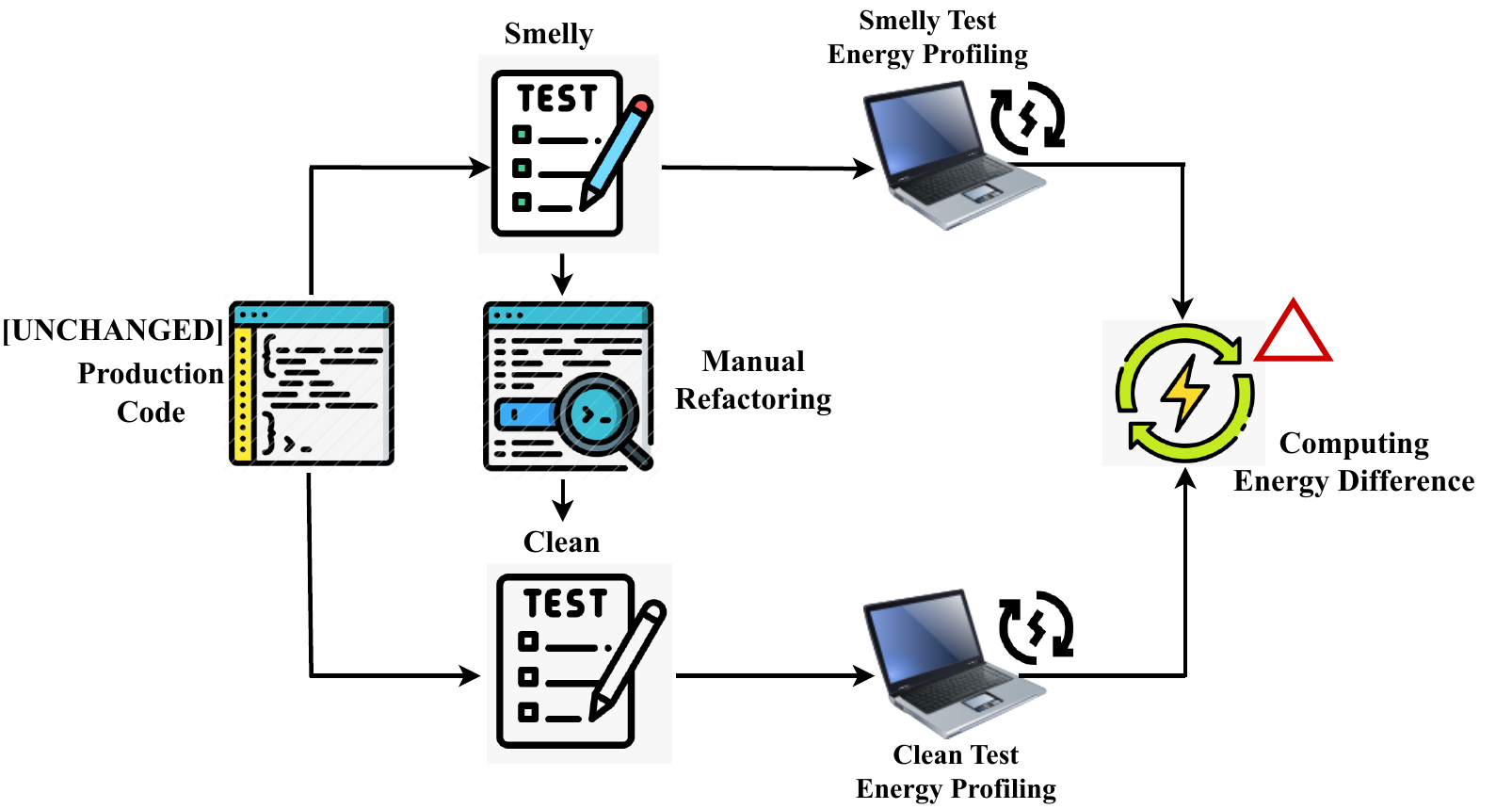}
\captionsetup{justification=centering, labelsep=colon, name=Figure}
\centering
\caption{Overview of Manual Refactoring of Test Smells and Energy Difference Computation}
\label{fig:refactor}
\end{figure}

\subsubsection{\textbf{Manual Refactoring of Test Smells}}
Automated test smell refactoring tools are not yet widely available. We conducted an exhaustive search of the existing literature and found only a few works proposing automated test smell refactoring approaches~\cite{santana2020raide,lambiase2020just,pizzini2022behavior,soares2022refactoring}. However, there is no executable tools available that we can readily utilize to do the automated test smell refactoring. We contacted the authors of these existing works and received several replies. None of the responses provide the proposed tool. Therefore, we took the refactoring strategies recommended by Soares et al.~\cite{soares2022refactoring} to refactor some of the test smell categories, namely Assertion Roulette (\texttt{AR}), Conditional Test Logic (\texttt{CTL}), Duplicate Assert (\texttt{DA}), Mystery Guest (\texttt{MG}), and Exception Handling (\texttt{EH}).
\par
We manually refactored the collected test cases by following the refactoring strategies proposed by Soares et al.~\cite{soares2022refactoring} while keeping the production code under test unchanged (Figure \ref{fig:refactor}). We performed partial refactoring for the ``\texttt{Smelly-66}'' test cases, focusing on addressing test smells that could be handled without modifying multiple test cases or introducing new ones. 
\par
\revise{
Figure~\ref{fig:smell_refactoring} demonstrates how we manually refactor a smelly test. Figure~\ref{fig:smelly_test} represents a smelly test case containing 4 instances of \textit{Assertion Roulette (AR)}, 3 instances of \texttt{Magic Number Test (MNT)}, 1 instance of \texttt{Lazy Test (LT)} and 1 instance of  \texttt{Eager Test (ET)} respectively, according to JNose test smells detection report. Since refactoring \texttt{ET} and \texttt{LT} requires to make changes on other test cases or introduce new test cases, we only refactored \texttt{AR} and \texttt{MNT} smells from this test case. To refactor \textit{AR}, we employed \texttt{JUnit5} \texttt{assertAll} feature as suggested by Soares et al.~\cite{soares2022refactoring}. This \texttt{assertAll} feature executes all assertions in parallel instead of the traditional sequential execution. Then, it reports any assertion failures in the group, which helps to identify the failed assertion and avoid the \texttt{AR} smells.}
\par
\revise{
For instance, in a test case where developers have multiple \texttt{assert} statements, a test case execution stops at a point when it encounters the first failed assertion. If there are subsequent failed assertions after the first one, those assertions will be ignored in the first run since the execution stops at the first assertion failure. If developers does not provide a explanation message for each assert statement, identifying the cause of the failure is not straightforward.}
\par
\revise{
However, with \texttt{JUnit5}-provided \texttt{assertAll} feature, developers can efficiently remove all \texttt{AR} smells. This is because \texttt{assertAll} executes all assertions in parallel and it reports all failed assertions together at the same time. Developers do not need the explanation in the assertions to identify the failed assertions. Instead, developers can identify and locate all failed assertions at once. Figure~\ref{fig:refac_test} shows the refactored version of the test case after removing the instance of \texttt{AR}, using the approach proposed by Soares et al.~\cite{soares2022refactoring} where they relied on the \texttt{assertAll} feature. In addition, to remove \texttt{MNT} smells from this test case, we defined two descriptive variables in the assertion statements.}
\par
\revise{
In this study, we performed complete refactoring for the test cases in the dataset ``\texttt{Smelly-AR-79}'' where we refactored all instances of the Assertion Roulette (\texttt{AR}) smells with \texttt{JUnit5} provide \texttt{assertAll} feature as suggested by Soares et al.~\cite{soares2022refactoring}.} Two of the authors collaborated to perform the test smell refactoring together. To ensure that the refactoring changes did not affect other test cases or existing features, we repeatedly executed the regression test suite provided by the project after refactoring. This step was taken to verify that the refactoring process did not introduce regressions or negatively impact the existing functionality. We also ensured that the code coverage did not change after refactoring. We posit that the combination of not introducing regression and unchanged code coverage ensures that the refactored test case is semantically similar to the original test case. \revise{The list of all smelly vs refactored testcases is available here.\footnote{\href{https://testcase-viewer.web.app/}{https://testcase-viewer.web.app/}}}
 
\begin{figure*}[b!]
\centering
\begin{subfigure}{.48\textwidth}
\begin{lstlisting}[style = mystyle, frame=none, label=lst:example,escapechar=\%, basicstyle={\scriptsize		 \ttfamily}]
95 : @Test
96 : public void testSingleQuoteMatcher() {
97 :   final StrMatcher matcher=StrMatcher.singleQuoteMatcher();
98 : %\CodeRedHilight%  assertSame(matcher, StrMatcher.singleQuoteMatcher());
99 : %\CodeRedHilight%  assertEquals(0,matcher.isMatch(BUFFER1, 10));
100: %\CodeRedHilight%  assertEquals(1,matcher.isMatch(BUFFER1, 11));
101: %\CodeRedHilight%  assertEquals(0,matcher.isMatch(BUFFER1, 12));
102: }
\end{lstlisting}
\caption{Smelly Testcase}
  \label{fig:smelly_test}
\end{subfigure}
\begin{subfigure}{.48\textwidth}
\begin{lstlisting}[style = mystyle, frame=none, label=lst:example, escapechar=\%,basicstyle={\scriptsize	 \ttfamily}]
95 : @Test
96 : public void testSingleQuoteMatcher() {
97 :   final StrMatcher matcher =StrMatcher.singleQuoteMatcher();
98 : %\CodeGreenHilight%  final int expected_0 = 0;
99 : %\CodeGreenHilight%  final int expected_1 = 1;
100: %\CodeGreenHilight%  assertAll(
101: %\CodeGreenHilight%   ()->assertSame(matcher,StrMatcher.singleQuoteMatcher()),
102: %\CodeGreenHilight%   ()->assertEquals(expected_0,matcher.isMatch(BUFFER1,10)),
103: %\CodeGreenHilight%   ()->assertEquals(expected_1,matcher.isMatch(BUFFER1,11)),
104: %\CodeGreenHilight%   ()->assertEquals(expected_0,matcher.isMatch(BUFFER1,12))
105: %\CodeGreenHilight% );
106: }
\end{lstlisting}
\caption{Refactored Testcase}
  \label{fig:refac_test}
\end{subfigure}
\caption{Test smells refactoring in \texttt{testSingleQuoteMatcher()} from \texttt{org.apache.commons.lang3.text.StrMatcherTest}.}
\label{fig:smell_refactoring}
\vskip-5mm
\end{figure*}

\subsubsection{\textbf{Energy Measurement}}
Finally, we have the smelly and refactored version of ``\texttt{Smelly-66}'' and ``\texttt{Smelly-AR-79}'' sampled test cases. We executed the refactored test case for five iterations of each test case with a 30-second cool-down period and measured the energy consumption. We describe the complete process of energy measurement in detail in Section \ref{sec:energy-measurement}. After the execution, we collected the energy consumption ($E$) for both smelly and refactored versions of the ``\texttt{Smelly-66}'' and ``\texttt{Smelly-AR-79}'' sampled test cases. Finally, we conducted Welch's t-test \cite{ruxton2006unequal,cohen2014applied} and Cohen's D~\cite{diener2010cohen} to measure the statistical significance and effect size of the differences in energy consumption between smelly and refactored test cases.

 \subsection{Developers Survey}
To validate our findings and understand developers' perceptions about test smell and its relationship to software energy consumption, we conducted an online survey with software developers. This survey was conducted following the guidelines and protocols approved by the Institutional Review Board (IRB). Our survey consists of 12 questions, including multiple-choice, ranking, and open-ended questions. The following subsections describe our survey design, participation selection, pilot survey, data collection, and analysis.
\par
\textbf{Survey Design:} We began by collecting demographic information from respondents (Q2-Q3) to understand their background and experience in software development and writing unit test cases. We then asked about their familiarity with test smells and their practice writing unit test cases (Q4-Q5). Additionally, we inquired whether they pay attention to test smells (Q6). Next, we investigated the impact of test smells on energy consumption during software testing (Q7-Q8). We identified specific test smells that had a higher energy consumption impact, and participants familiar with these were asked to rate the test smells based on their perceived impact severity on energy consumption (Q8-Q9). Additionally, we inquired whether participants' organizations provide guidelines for monitoring energy consumption during software testing. Participants were asked to mention any tools or services they use to monitor energy consumption during testing (Q10-Q12). A text box option was also provided for respondents if they wanted to share the reason behind their choice. A complete list of questions for this survey is provided on the companion website \cite{supply}.
\par
\textbf{Participants Selection:} For our survey, we targeted the software developers from the 12 subject Apache projects in our study. To develop a list of survey participants, we mined a list of unique email addresses of contributors from the version control systems. In total, we collected 490 individual email addresses and recognized them as our potential participants. We utilized this email list to send the survey invitation to these developers.
\par
\textbf{Pilot Survey:} To review the survey's validity, we asked Software Engineering professors and graduate students (two professors and two Ph.D. students) with experience in software development, writing unit test cases, and survey design. To enhance the clarity of the questions, we performed several iterations of the survey and rephrased and reorganized some questions according to their feedback. Considering software developers' hectic schedules, we emphasized the time required to complete the survey. We ensure that participants can complete the survey in 8 to 10 minutes. The pilot survey aimed only to improve the questions, and the responses are not included in the reported results.\\
\par
\textbf{Data Collection:} To distribute our survey, we used Qualtrices \cite{Qualtrics} as a design and distribution platform. We emailed 490 developers from 12 Apache projects, following our organizational guidelines (with the approved University IRB protocol). To maximize survey participation, we followed the guidelines and best practices described by Smith et al.~\cite{smith2013improving}, such as allowing respondents to remain anonymous and sending personalized invitations. After publishing the survey, the survey was kept open for two weeks in total, and meanwhile, we also sent a reminder email at the end of the first week. Within these two weeks, we received completed responses from 62 participants. Overall, we got a response rate of 12.7\%, consistent with the prior studies conducted in the software engineering areas \cite{smith2013improving,lo2015practitioners}.
The software development experience of our respondents varies from 1 year to more than 20 years, and 80.6\% of them have over two years of software testing experience.
\par
\textbf{Data Analysis:} During the data analysis process, we consider the majority vote from developers as the final overall rating for a specific item. For example, when ranking test smell types in terms of the severity of their impacts on energy consumption, we determine the final ranking based on the majority consensus among the developers' responses. This approach ensures that the final results represent the collective perspective of the surveyed developers. 

 \section{Results}
\label{sec:results}
In this section, we present the results of our study from two complementary perspectives: \textit{Software} and \textit{Developers'}.

\subsection{Software Perspective}
\label{subsec:softwareresults}
We start this section by reporting our findings on how software energy consumption is associated with the presence of test smells.

\begin{figure}[t]
\centering
\includegraphics[width=0.80\linewidth]{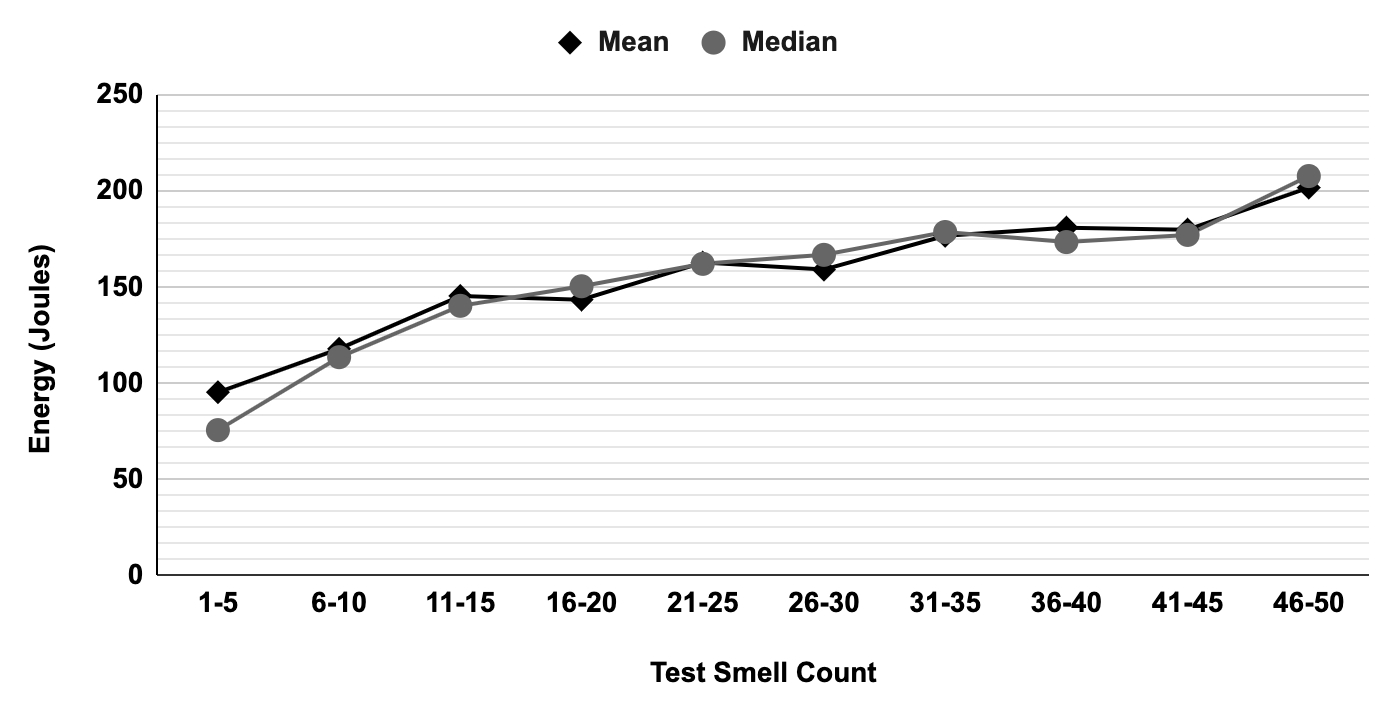}{}
\captionsetup{justification=centering, labelsep=colon, name=Figure}
\caption{ Mean ($E_{LOC_{smell}}$) and Median ($E_{LOC_{smell}}$) energy consumption with Test Smells Count ($SC$) in Group size 5. Energy Unit Measured in Joules.}
\label{fig:rq1-group-analysis}
\end{figure}
\subsubsection{{\textbf{RQ1 [Test Smells vs. Energy]} How do smelly tests, in general, impact energy consumption?}}
\label{sec:rq1-answer}
To answer this research question, we first investigate if the smelly part of a test case consumes more energy than its clean part. We calculate the energy consumption for smelly lines of code ($E_{LOC_{(smell)}}$) and clean line of code ($E_{LOC_{(clean)}}$) following Equations \ref{eq:1} and \ref{eq:2} mentioned in Section \ref{sec:group-analysis} for all smelly test cases in our selected projects. The values $E_{LOC_{(smell)}}$ and $E_{LOC_{(clean)}}$ represent an estimate of the energy consumption for smelly and clean test codes in a test case. We found that the mean value for the smelly test code is 109.93 Joule, 
which is greater than that of the clean test code, which is 88.27 Joule. The median value also follows the same trend with a higher energy consumption value for the smelly test code (102.61  Joule) compared to the clean test code (91.66  Joule). Further, to check whether the difference between the energy consumption of the smelly test code and the clean test code across all 12 projects is statistically significant, we performed Welch's t-test \cite{ruxton2006unequal} after checking for the normality assumption of the data using Shapiro-Wilk's test ~\cite{yazici2007comparison}. We used Cohen's D \cite{diener2010cohen} to measure the effect size. Our results show that the difference is statistically significant (Welch's t-test, \textit{p-value} $<$6.27e-120, Cohen's D (0.37, small)), indicating that in a test case, the smelly lines consume more energy than the clean lines of code. \revise{To demonstrate how the smelly part of the code consumes more energy, Figure~\ref{fig:smelly_part} depicts an example smelly test case containing \texttt{AR}, \texttt{MNT}, \textit{LT} and \textit{ET} smells. The \textit{AR} , \textit{MNT} and \textit{ET} smells are confined within the the assertion statements, which would consume more energy. 
Thus, the total energy consumption of this test are heavily influenced by the highlighted smelly lines.}
\par
In addition, we also examine if more test smell instances in a smelly test case would cause it to consume more energy. To do so, we grouped all test cases based on the count of total number of smell instances it contains ($SC$). To prevent outliers from skewing our results, we removed 79 test cases having more than 50 test smell instances from our analysis (3 standard deviations away from the mean~\cite{ilyas2019data}). This process left us with 7,748 test cases containing at least one but no more than 50 test smell instances. Next, we group test cases based on the number of test smell instances each test case has. Figure \ref{fig:rq1-group-analysis} shows the mean and median energy consumption in terms of smelly line of code ($E_{LOC_{(smell)}}$) values for the group size of 5. As we can see, the energy consumption of smelly line of code ($E_{LOC_{(smell)}}$) grows with the increasing number of test smells. For group sizes 10 and 25, we found the same pattern. In general, it exhibits that the presence of more test smell instances increases the energy consumption for a test case, which implies test smell, in general, impacts energy consumption in software testing. To validate the statistical significance of the difference between groups, we also conducted Welch's t-test and Cohen's D between contiguous groups (i.e., group 6-10 with group 11-15, group 11-15 with group 16-20, etc.). We provide the complete group analysis results in our replication package~\cite{supply}.

\begin{tcolorbox}[width=0.48\textwidth, colframe=black, colback=anti-flashwhite!30, boxsep=1mm, arc=1.5mm]
\textbf{Observation 1: The smelly part of a test case consumes more energy than its non-smelly part}. 
\end{tcolorbox}

\begin{figure}[t]
\centering
\begin{lstlisting}[style = mystyle, frame=none, label=lst:example,escapechar=\%, basicstyle={\scriptsize		 \ttfamily}]
232: @Test
233: public void testSimpleStreamMap() {
234:     final List < String > input = Arrays.asList("1", "2", "3", "4", "5", "6");
235:     final List < Integer > output = Failable.stream(input)
        .map(Integer::valueOf)
        .collect(Collectors.toList());
236:%\CodeRedHilight%     assertEquals(6, output.size());
237:     for (int i = 0; i < 6; i++) {
238:%\CodeRedHilight%         assertEquals(i + 1, output.get(i).intValue());
239:     }
240: }

\end{lstlisting}
\caption{An smelly test case \texttt{testSimpleStreamMap()} from \texttt{org.apache.commons.lang3.stream.StreamsTest}}
  \label{fig:smelly_part}
\end{figure}

\subsubsection{ {\textbf{RQ2 [Test Smell Types vs. Energy]}~How does each test smell type impact energy consumption?}}
\label{sec:rq2-result}
We next seek which test smell types have a more severe impact on energy consumption than others. As discussed in Section \ref{sec:correlation-analysis}, it is not possible to collect the actual one-to-one
mapping of energy consumption and each type of smell. Therefore, we computed a normalized energy consumption
value $E_{(N)}=(E/SC)$ as an indicator of test smell's impact on energy consumption (shown in equation 3). In a test case, the
normalized value $E_{(N)}$ helps us to create a one-to-one mapping between $E_{(N)}$ and the each type of
test smells such as $testcase \rightarrow \{ E_{(N)},s_{1}, s_{2},s_{3}...s_{21}\}$.
\par
With this mapping, we then performed Kendall's Tau ($\tau_{b}$) correlation analysis between $E_{(N)}$ and each type of smell (e.g., $s_{1}, s_{2},s_{3}...s_{21}$). Table \ref{tab:rq2-corelation} demonstrates our correlation results with the top 10 types of smells. Since we perform multiple-statistical tests, we applied Bonferroni correction to adjust $P$ values~\cite{napierala2012bonferroni}, which gives an adjusted $\alpha = 0.002$. As we can see, the correlation values Kendall ($\tau_{b}$) are statistically significant (i.g., \textit{p-value $<$ 0.002}) for all these test smells. Assertion Roulette (\texttt{AR}), Lazy Test (\texttt{LT}), Eager Test (\texttt{ET}), and Magic Number Test (\texttt{MNT}) are the smells with strong ($\tau_{b} > 0.30$) correlation with energy consumption. On the other hand, Dependent Test (\texttt{DT}), Unknown Test (\texttt{UT}), and Verbose Test (\texttt{VT}) are moderately ($\tau_{b} > 0.20$) correlated with energy consumption. Sensitive Equality (\texttt{SE}) and Conditional Test Logic (\texttt{CLT}) show a weak correlation. For the remaining test smells types, we found a very weak correlation ($\tau_{b} < 0.10$).
\par
\begin{tcolorbox}[width=0.48\textwidth, colframe=black, colback=anti-flashwhite!30, boxsep=1mm, arc=1.5mm]
\textbf{Observation 2: Assertion Roulette (\texttt{AT}), Lazy Test (\texttt{LT}) and Eager Test (\texttt{ET}) test smells are strongly associated with energy consumption.
}
\end{tcolorbox}

\begin{table*}[b]
\centering
\caption{Top 10 Test Smells based on the correlation between Energy/Smell ($E_{N}$) vs Each type of test smell}
\label{tab:rq2-corelation}
\begin{tabular}{|l|l|l|l|}
\hline
\textbf{Test Smell} & \textbf{Rank} & \textbf{Kendall} ($\tau_{b}$) & \textbf{\textit{p-value}}    \\ \hline
\texttt{AR (Assertion Roulette) }        & 1    & 0.615   & $<$ 0.002 \\ \hline
\texttt{LT (Lazy Test)}         & 2    & 0.449   & $<$ 0.002 \\ \hline
\texttt{ET (Eager Test)}         & 3    & 0.432   & $<$ 0.002 \\ \hline
\texttt{MNT (Magic Number Test)}        & 4    & 0.385   & $<$ 0.002 \\ \hline
\texttt{DA (Duplicate Assert)}         & 5    & 0.290   & $<$ 0.002 \\ \hline
\texttt{UT (Unknown Test)}         & 6    & 0.247   & $<$ 0.002 \\ \hline
\texttt{VT (Verbose Test)}         & 7    & 0.246   & $<$ 0.002 \\ \hline
\texttt{SE (Sensitive Equality)}         & 8    & 0.177   & $<$ 0.002 \\ \hline
\texttt{CLT (Conditional Test Logic)}        & 9    & 0.172   & $<$ 0.002 \\ \hline
\texttt{RA (Redundant Assertion)}         & 10   & 0.072   & $<$ 0.002 \\ \hline
\end{tabular}
\end{table*}

\subsubsection{ {\textbf{Case Study [Smelly/refactored Test vs. Energy]}}}
To complement our results from RQ1 and RQ2, we conducted a case study on test cases with and without test smells (See Section ~\ref{casestudy}). For ``\texttt{Smelly-66}'', a significant difference (Welch's t-test, \textit{p-value} $<$6.727e-46, Cohen's D (3.910, large)) was found between the energy consumption ($E$) of smelly test cases (204.922 Joule) and partially refactored test cases (184.974 Joule). Similarly, for ``\texttt{Smelly-AR-79}'', we also found a significant difference (Welch's t-test, \textit{p-value} $<$4.508e-05, Cohen's D (0.672, medium)) between the energy consumption ($E$) of smelly test cases (204.048 Joules) and clean test cases (185.017 Joules). Our results show that the total energy consumption decreased significantly after removing test smells for both partial and complete refactoring. This indicates that test smells incur more energy consumption in software testing.

\subsection{Developers' Perspective}
Here, we explain our results regarding the developer's practices and perception of test smells.

\subsubsection{{\textbf{RQ3 [Developers’ Awareness]}
\label{sec:rq3-answer}
Are developers aware of the impact of test smells on energy consumption?}}
Only 29.4\% of our survey respondents who know about test smells expressed that they are confident that the presence of test smells in test cases has an impact on energy consumption, while 70.6\% answered ``Maybe" or ``No." This indicates that most developers are not fully aware of the test smell's impact on energy consumption, which may contribute to the introduction of test smells during software testing.

\begin{figure}[t]
    \centering
    \includegraphics[width=0.80\linewidth]{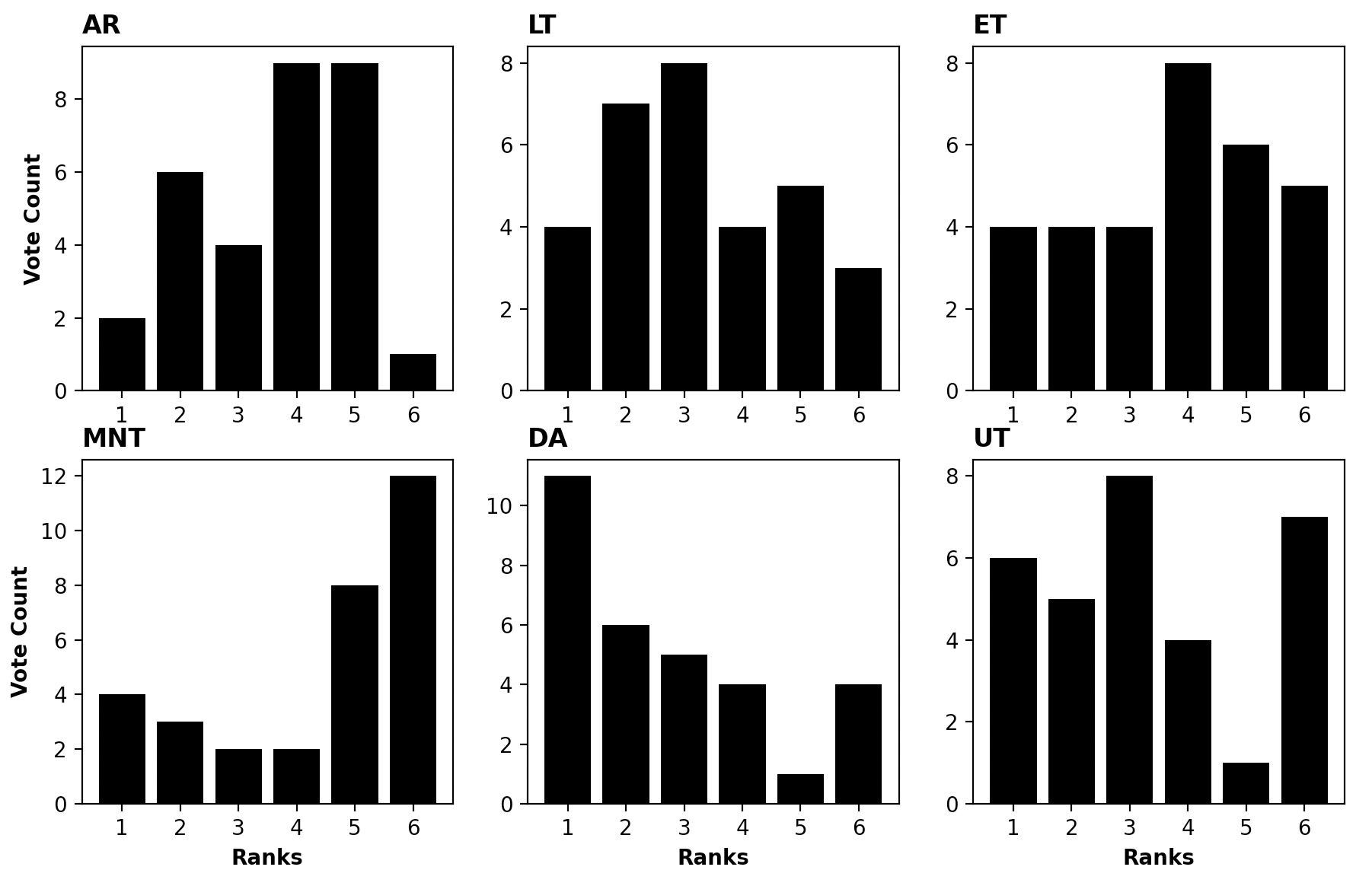}
    \captionsetup{justification=centering, labelsep=colon, name=Figure}
    \caption{Rankings of the Severity of Test Smell Types' Impact on Energy Consumption from Survey Respondents}
    \label{fig:survey_ranking}
\end{figure}

We asked the survey participants to rank the test smell types we found in RQ2 based on their perceived severity of impacts on energy consumption. We show survey respondents' provided rankings in Figure \ref{fig:survey_ranking}. We could see that the rankings are widely spread among developers, which indicates that developers have conflicting opinions regarding the severity of different test smell types' impacts on energy consumption. We took the ranking that is voted by most respondents as the overall ranking from developers for a specific test smell type. We list our results in Table \ref{tab:RQ4}. We found multiple ranking mismatches between our empirical analysis and our survey respondents, such as Duplicate Assert (\texttt{DT}), Unknown Test (\texttt{UT}), and Assertion Roulette (\texttt{AR}). This, again, shows that developers have a limited understanding of the test smell's impact on energy consumption.

\begin{tcolorbox}[width=0.48\textwidth, colframe=black, colback=anti-flashwhite!30, boxsep=1mm, arc=1.5mm]
\textbf{Observation 3: Most developers are unaware of the impact of test smells on energy consumption.}
\end{tcolorbox}

\begin{tcolorbox}[width=0.48\textwidth, colframe=black, colback=anti-flashwhite!30, boxsep=1mm, arc=1.5mm]
\textbf{Observation 4: There is a mismatch in the ranking of the impact severity between developers' perception and our empirical analysis.}
\end{tcolorbox}

\begin{table}[t]
\centering
\caption{Comparison of Rankings Based on Empirical Analysis and Developers' Response}
\label{tab:RQ4}
\begin{tabular}{|l|p{1.5cm}|p{1.5cm}|}
\hline
\textbf{Test Smell} & \textbf{Rank (Empirical Analysis)} & \textbf{Rank (Developers)} \\ \hline
\texttt{AR (Assertion Roulette)  }                      & 1                        & 4                          \\ \hline
\texttt{LT (Lazy Test)}                       & 2                        & 3                          \\ \hline
\texttt{ET (Eager Test)}                        & 3                        & 4                          \\ \hline
\texttt{MNT (Magic Number Test)}                      & 4                        & 6                          \\ \hline
\texttt{DA (Duplicate Assert)}                       & 5                        & 1                          \\ \hline
\texttt{UT (Unknown Test)}                        & 6                        & 3                          \\ \hline
\end{tabular}
\end{table}

 \subsubsection{{\textbf{RQ4 [Provenance of Test Smell]}
\label{sec:rq4-answer}
What are the underlying reasons developers introduce test smells that could cause additional energy consumption?}}
First, 45.2\% of our survey participants do not know about test smells, so they may unknowingly introduce test smells when writing and updating test cases. Then, 5.9\% of our survey respondents who know about test smells do not pay attention to test smells when writing test cases. One of them explicitly mentioned that \textit{``My organization doesn't have any policy/requirement regarding test smells"} and \textit{``I don't care about test smells in test cases"[S-18]}. In addition, 61.8\% of the developers replied that their organizations do not follow any guidelines regarding monitoring energy consumption during software testing, while 76.5\% do not know or use any tools for monitoring energy consumption in the testing phase. The complete results of our survey is provided in our replication package~\cite{supply}. From our survey results, we could see that even well-established organization like Apache lacks the proper guidelines for test smells and energy consumption monitoring in regular software testing, which might be one of the main reasons developers introduced energy-hungry test smells.

\begin{tcolorbox}[width=0.48\textwidth, colframe=black, colback=anti-flashwhite!30, boxsep=1mm, arc=1.5mm]
\textbf{Observation 5: Lack of guidelines, tools, and incentives are probable reasons developers introduce energy-hungry test smells.}
\end{tcolorbox}

 \section{Discussion}
\label{sec:discussion}
The results of our study reveal that test smells, in general, have a negative impact on energy consumption during software testing. This finding complements previous research on the impact of test smells on various other aspects of software quality. Our study provides valuable evidence for developers, highlighting that the presence of test smells can lead to a significant amount of energy overhead, especially in large organizations where millions of test cases are executed daily.
\par
To address this issue and promote energy-efficient software development practices, tool builders should consider providing Just In Time automated tools and IDE plugins. One surprising finding was the lack of automated test smell refactoring tools. As explained in Section~\ref{casestudy}, even though there is prior work on test smell refactoring, no tool is available to refactor test smells off the shelf. Providing tools that can help developers by offering real-time feedback and suggestions for refactoring, developers can make more informed decisions to create cleaner and more energy-efficient test cases, ultimately leading to improved software quality and reduced energy consumption in the testing process.
\par
Our analysis also revealed that the top three test smells with the highest association with energy consumption are Assertion Roulette (\texttt{AR}), Lazy Test (\texttt{LT}), and Eager Test (\texttt{ET}). To understand why \texttt{AR} is particularly energy-intensive, we manually inspected samples of \texttt{AR} test smell instances. Our investigation showed that most test cases containing \texttt{AR} have multiple assertion statements without any explanation, and all of these statements are considered smelly. In contrast, clean test cases usually contain only one assertion statement, which is more focused and purposeful. From an execution perspective, multiple assertion statements in Assertion Roulette (\texttt{AR}) test cases require more energy during testing than clean test cases with only one assertion statement. The repeated execution of multiple statements adds to the overall energy consumption, making test cases with Assertion Roulette (\texttt{AR}) more energy-intensive. We believe more research is needed to systematically investigate why these test smell types are the most energy-hungry ones.
\par
Our analysis found a mismatch between the developer's perception and our empirical analysis result regarding the severity of test smell types' impacts on energy consumption. For example, the most energy-hungry test smell type recognized by developers is Duplicate Assert (\texttt{DA}) while it only ranks 5 in our empirical analysis. One possible explanation could be that developers assumed multiple unnecessary assertions in a test case could contribute to more execution time and energy consumption. Such mismatches between perception and reality have been observed in numerous other instances where long-held beliefs proved incorrect or outdated when actual evidence was collected through empirical analysis~\cite{DBLP:conf/icse/Devanbu0B16,DBLP:journals/ese/MenziesNSL17}. Since the developers do not have tools for monitoring energy consumption during software testing, tool builders should create such tools that can be seamlessly integrated into the existing development workflow.
\par
In terms of introducing energy-hungry test smells, bot committers (i.e., automated tools) are responsible for a comparable number of test smell instances to core and non-core developers. Combined with observations from~\cite{DBLP:conf/icse/PalombaNPOL16,DBLP:journals/jss/GranoPNLG19,DBLP:conf/sbes/VirginioMSSCM20}, we believe that researchers should investigate ways to consider energy consumption as a factor while generating test cases automatically. Another interesting observation is that non-core developers are responsible for a similar amount of test smell instances as core developers, even though they contribute much less than core developers. Further investigation in the future is required to understand the underlying reason for this.

 \section{Threats To Validity}
\label{sec:ttv}
In our study, we have tried to eliminate bias and the effects of random noise. However, a few biases are unavoidable, and our mitigation strategies may not have been effective for them.

\textbf{Bias due to confounding factors:} The potential confounding effect of the production code's complexity or LOC is a concern when studying the correlation between test smells and energy consumption. However, the case study conducted in Section ~\ref{casestudy} effectively mitigated this bias by ensuring that the production code remained unchanged across test cases with and without test smells as shown in Figure~\ref{fig:refactor}. By keeping the production code unchanged, we isolate the impact of test smells on energy consumption and ensure that any observed changes in energy consumption are attributed only to the presence or absence of test smells in the test cases. 

\textbf{Bias due to sampling:} Energy consumption studies have been done in many domain-specific software (e.g., mobile apps) and programming languages. However, our work is specific only to test smells and their impact on energy consumption. The projects we have used in our study includes 12 Apache Java projects. We picked these projects from the Apache Software Foundation. Besides, we surveyed the developers who contributed to these 12 Apache projects. The responses of these developers may not represent all developers in other open-source projects and therefore, our findings may not generalize to all open-source projects. However, the experimental phases of our study is still applicable to other programming language projects and domain specific software.
\par
We mined in total 13,103 test cases. We detect the test case based on the presence of \texttt{@Test} annotation. However, developers may write test cases without the \texttt{@Test} annotation by extending the JUnit Test class. We followed the developer's best practices and guidelines mentioned in JUnit  documentation to identify test cases. However, it is also possible that other libraries are used to write unit test cases. In our study, we didn't consider test cases written using other libraries. 
\par
It is possible that our sampled test cases for manual refactoring are not representative of all test cases. However, to mitigate this threat, we utilized a 90\% confidence interval and a 10\% margin of error to calculate the sample size and used stratified random sampling. This statistical approach should mitigate the mentioned bias.
\par
\textbf{Bias due to energy metrics:} Although different code lines (e.g., simple statement, conditional or branch statement) probably consume energy very differently, during our analysis of test smells impact, we normalized the energy consumption values according to Equation ~\ref{eq:1} and~\ref{eq:2}. Since test smells can appear both in single and in multiple lines, we believe our normalization represented amortized energy cost of each line of code execution. Using normalized values other than Equation~\ref{eq:1} and~\ref{eq:2} will alter our findings. \revise{In addition, the results of the energy measurement by the energy profiler adopted in our study may not generalize to other hardware configurations.}
\par
\textbf{Bias due to tools used:} In our study, we utilized a test smell detection tool and an energy profiler. Our analysis depended on these tool-generated outputs. Therefore, any errors in these tools may affect findings. To minimize the risk, we used tools that were validated by prior research. We used JNose to detect test smells. JNose can detect 21 types of test smells. However, there could be other test smell types that JNose can not detect.
\par
We employed Intel PowerLog as an energy profiler. Although it reports energy consumption at the software level, some environmental factors, such as room temperature and outage in electricity, can affect its reported energy consumption result. We followed the procedure used in other work; we kept our experimental laptops at normal room temperature and placed them at a reasonable distance from each other to avoid heat transmission. We also ensured a cool-down period after every test case execution to avoid any impact of heating of the laptop itself on the results. \section{Conclusion}
\label{sec:conclusion}
Our ultimate goal is to help catalyze advances in energy-efficient software testing and this paper takes the first step towards that by shedding light on the current state of affairs. We presented the results of our empirical study aimed at understanding the state of energy consumption occurring due to test smells. Overall, our analysis reveals that the smelly part of a test case generally consumes more energy than its clean part.
\par
Also, not all test smell types are equally energy-hungry. Our analysis revealed that Assertion Roulette (\texttt{AR}), Lazy Test (\texttt{LT}), and Eager Test (\texttt{ET}), tend to consume more energy compared to other smell types. Moreover, our test smell refactoring results indicate that smelly tests consume more energy than their clean counterparts.
\par
Our findings highlight the need for increased developer awareness regarding the impact of test smells on energy consumption and opportunities for researchers and tool builders to address the lack of tools and guidelines. \revise{As a future work, we plan to extend our study to a broader context of software design issues on energy consumption including smells in production code (i.e., code smells). For example, we can compare the energy consumption savings by removing only production code smells under test and only the test smells to provide more insights into sustainable and energy-efficient software maintenance practice.} The research artifacts for this study are publicly available at the companion website~\cite{supply}. 

\balance
\bibliographystyle{elsarticle-num}
\bibliography{references}

\end{document}